\documentclass[]{aastex}
\usepackage{aas_macros}
\usepackage{natbib}

\begin{document}

\title{``Retired'' Planet Hosts: Not So Massive, Maybe Just Portly After Lunch}

\author{James P. Lloyd}
\affil{Department of Astronomy, Cornell University, Ithaca NY}

\date{}

\begin{abstract}

Studies of the planet abundance as a function of stellar mass have suggested a strong increase in the 
frequency of planet occurrence around stars more massive than $1.5~M_\odot$, and that such stars are deficit in short period planets.  These planet searches have relied on giant stars  for a sample of high mass stars, which are hostile to precision Doppler measurements due to rotation and activity while on the main sequence.  This paper considers the observed $v\sin i$ and observationally inferred mass for exoplanet hosting giants with the  $v\sin i$ of 
distribution of field stars, which show discrepancies that can be explained by erroneous mass determinations of some exoplanet host stars.  By comparison with an expected mass distribution constructed from integrating isochrones, it is shown that the exoplanet hosts are inconsistent with a population of massive stars. These stars are more likely to have originated from a main sequence population of late F/early G dwarfs with mass $1.0$--$1.2~M_\odot$, only slightly more more massive than the typical FGK dwarfs with Doppler detected planets.  The deficit of short period planets is most likely explained by tidal capture.  The planet abundance difference requires either a steeper increase in planet frequency with mass than previously thought or a high rate of false positives due to signals of stellar origin.  The measurement of photospheric Carbon isotope ratios is suggested as a method to discriminate whether this sample of giant stars is significantly more massive than the population of FGK dwarfs with Doppler detected planets.

\end{abstract}


\maketitle

\section{Introduction}

The number of exoplanets has reached the point that statistical studies may provide clues to planet
formation by revealing correlations such as the planet-metallicity correlation \citep{Fischer:2005mz}.  
While planet formation is expected to be influenced by stellar mass,
planet discoveries have been biased toward FGK dwarfs.  Stars outside this range are the subject of intense interest, but M dwarfs are intrinsically faint and challenging for RV surveys \citep{Muirhead:2011dq} and higher mass main sequence stars have broad lines and activity that degrades RV precision \citep{Lagrange:2009fr}.

Recent progress has been made studying ``retired A stars'' that are more amenable to Doppler RV, having evolved and developed slowly rotating cool atmospheres.  This population has shown two distinct differences from the planets orbiting FGK dwarfs: an increased planet abundance and paucity of planets in short period orbits \citep{Johnson:2007kx,Hekker:2008rt,Johnson:2010uq,Bowler:2010yq}.  
The distribution of planets orbiting massive stars is central to predicting the yield of imaging surveys  \citep{Crepp:2011lr,McBride:2011qy,Beichman:2010vn} which benefit from the conditions of youth, wide separation and massive planets that are most likely around massive stars.

\section{Rotational evolution}

Late stage accretion in the formation of stars is angular-momentum limited, giving rise to ubiquitous rapid rotation of young stars.  After dispersal of the protostellar disk, the evolution of stellar rotation includes at least two components.  The star can lose angular momentum to a magnetized stellar wind, as is responsible for the spin-down of FGK stars. The lack of this mechanism in massive stars on the main sequence renders them difficult for radial velocity planet searches.   Conservation of angular momentum dictates that the angular velocity decrease as a star expands due to evolution.  If the star has planets, there is a third possibility: the transfer of angular momentum to or from the planets.  

There are applicable theories to the tidal evolution of planetary systems, and theoritical work is investigating 
 the question of tidal evolution and the deficit of short period planets orbiting intermediate mass stars \citep{Brown:2011ai,Carlberg:2011sp,Kunitomo:2011tg,Madappatt:2011dp,Matsumura:2010hc}.  However, there is a large and longstanding uncertainty in the effective tidal dissipation parameters $Q'$ in both star \citep{Penev:2011ij} and planet \citep{Goldreich:1977th}.  In the case of evolving subgiant stars, the problem of estimating $Q$ would seem acute, as the star is developing a convective envelope, and the torque occurs at the convective-radiative boundary \citep{Goldreich:1989bs}.  Regardless of details, the evolution is a steep function of stellar radius: tidal evolution rate increases as $R_*^{10}$ \citep{Hansen:2010fy}.  Since the radius of a star increases by two orders of magnitude through the giant phase, it should be expected that regardless of tidal dissipation mechanisms, if tidal interaction with the star in any way shapes the planetary system evolution, it must impact the rotation the star.

The direction of angular momentum transfer depends on whether the planet orbits inside or outside of co-rotation.   In the case of a star rotating faster than the orbital period, the angular momentum must transfer to the planet until the star rotates synchronously with the planet's orbit.  Tidal evolution can only decrease the semimajor axis if the star can absorb the angular momentum.   Planets are, as the relic of the angular momentum supported protostellar disk, typically the largest reservoir of angular momentum (only 4\% of the angular momentum of the solar system resides in the sun).  If the tidal evolution ends with the planet merging with the star then the angular momentum must all be deposited in the star.  

In the case of the intermediate mass stars, there are two situations that might show clear observational evidence for tidal evolution.   Since intermediate mass stars are born with near-breakup rotation and lose little of it on the main sequence, the first interaction between the evolving star and planet would be for star to deposit most of its  angular momentum in the planet.   If a star accretes a planet, in the absence of the transfer of angular momentum to a third body or loss through a wind, the angular momentum of that planet will substantially increase the total angular momentum of the star.  Since intermediate mass stars retain a large fraction their initial angular momentum through the main sequence, a subgiant that showed angular momentum larger than the breakup angular momentum of the zero age main sequence (ZAMS) star from which that subgiant evolved would be a compelling case for a star that has absorbed a planet.   Motivated by these considerations, this paper begins by investigating the observed rotation of evolved stars with planets.  

\section{Rotation of Planet Host Stars}

I have examined the rotation (as determined by $v\sin i$) of evolved planet host stars with $\log~g<4$ from the Exoplanet Orbit Database \citep{Wright:2010rt}.  A comparison sample of field stars with similar mass and $\log g$ can be constructed from the the catalog of Hipparcos-determined physical parameters of \citet{Allende-Prieto:1999yq} and the radial velocity catalog of \citet{Glebocki:2000vn}.

\begin{figure}[htbp]
\begin{center}
\includegraphics[width=5in]{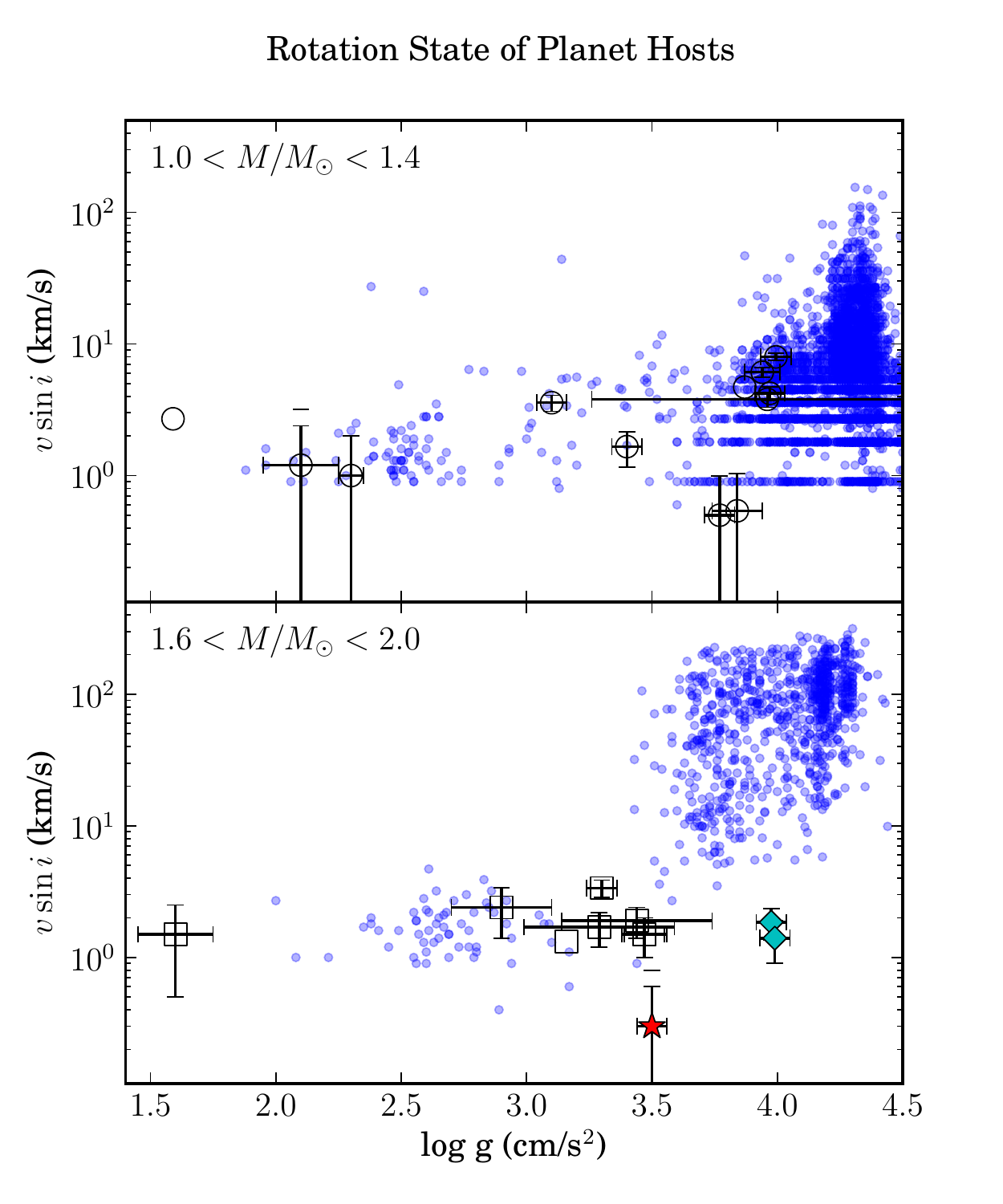}
\caption{Top panel: $v\sin i$ and $\log g$ for exoplanet hosts with masses $1.0$--$1.4~M_\odot$ (open circles) and a comparison sample of field stars with parameters determined in \citet{Allende-Prieto:1999yq} and $v \sin i$ from \citet{Glebocki:2000vn}.  Lower panel: exoplanet hosts with masses $1.6$--$2.0~M_\odot$ .  Cyan diamonds mark HD 190228, HD 154857 and the red star marks HD 102956, which are discussed in Section~\ref{sec:specificstars}.  The exoplanet host star $v \sin i$, masses and uncertainties were those reported in the Exoplanet Orbit Database \citep{Wright:2010rt}.   Uncertainties in $\log g$ reported by \citet{Allende-Prieto:1999yq} are in general $0.05-0.1$ dex.}
\label{fig:rotationstate} 
\end{center}
\end{figure}

A sub-sample in this comparison of planet hosts with masses between 1.6 and 2.0 $M_\odot$ (shown in the lower panel of Figure~\ref{fig:rotationstate}) shows some discrepancies with comparison field stars.  Three planet hosts are more slowly rotating than comparison stars, and the planet hosts occupy a range of $\log g$ that is sparsely populated.  Seven of the eleven planet hosts with $1.6<M/M_\odot<2.0$ have $3.0<\log~g<3.5$, but only seven slowly rotating stars with $1.6<M/M_\odot<2.0$ and $3.0<\log~g<3.5$ are within the 100 pc volume of the \citet{Allende-Prieto:1999yq} catalog. 

\subsection{Host Star Masses}

\begin{figure}[htbp]
\begin{center}
\includegraphics[width=5in]{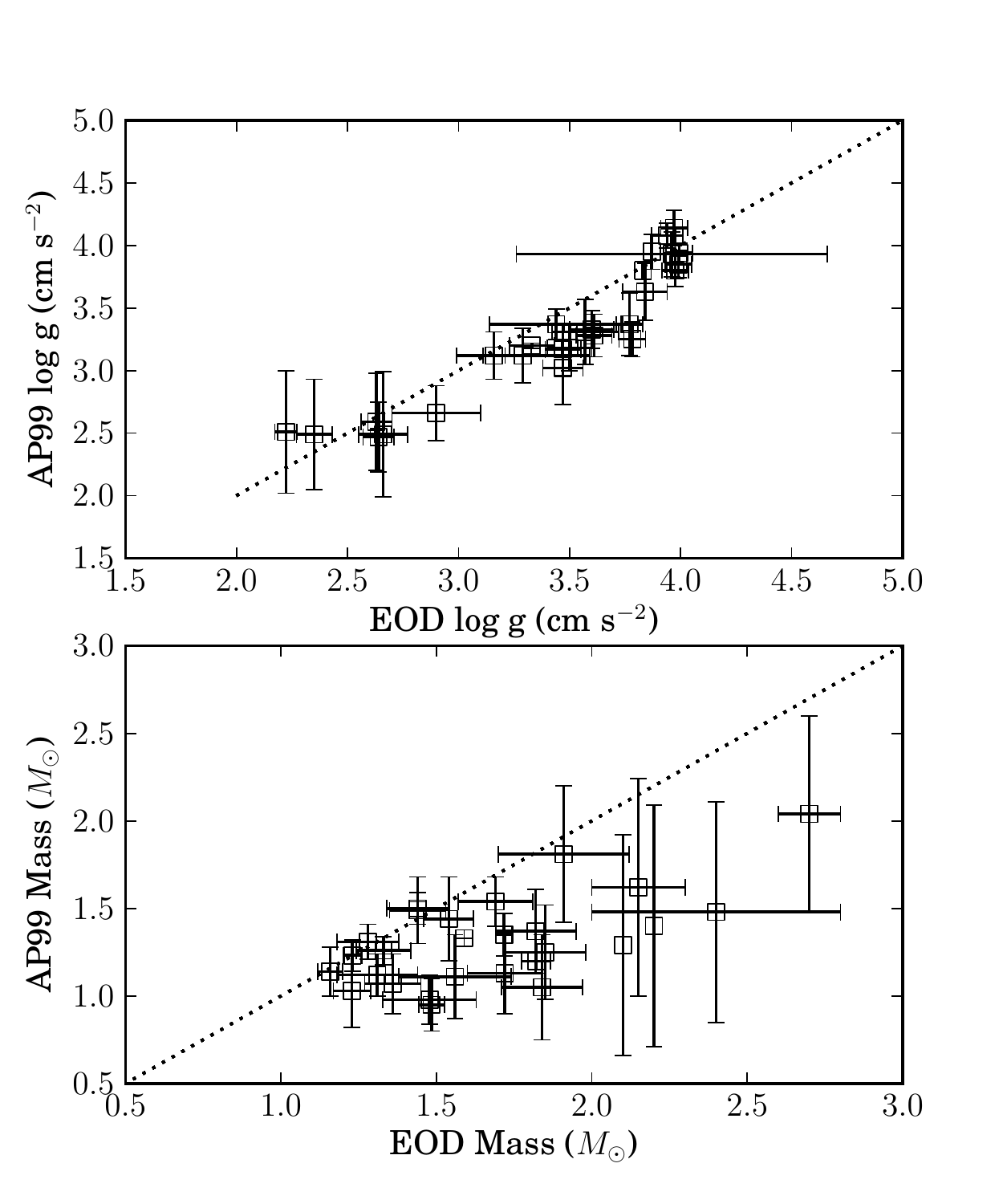}
\caption{Physical parameters of the 25 evolved stars ($\log~g>4$) in common between the Exoplanet Orbit Database (EOD) and \citet{Allende-Prieto:1999yq} (AP99).  Gravity (upper panel) determined by the differing methods agree reasonably well, with half of within $1\sigma$ and all but one within $2\sigma$.  Masses (lower panel) disagree, with the spectroscopic EOD masses being larger than the AP99 photometric masses.}
\label{fig:EODvsAP99}  
\end{center}
\end{figure}

\label{sec:specificstars}

\begin{figure}[htbp]
\begin{center}
\includegraphics[width=5in]{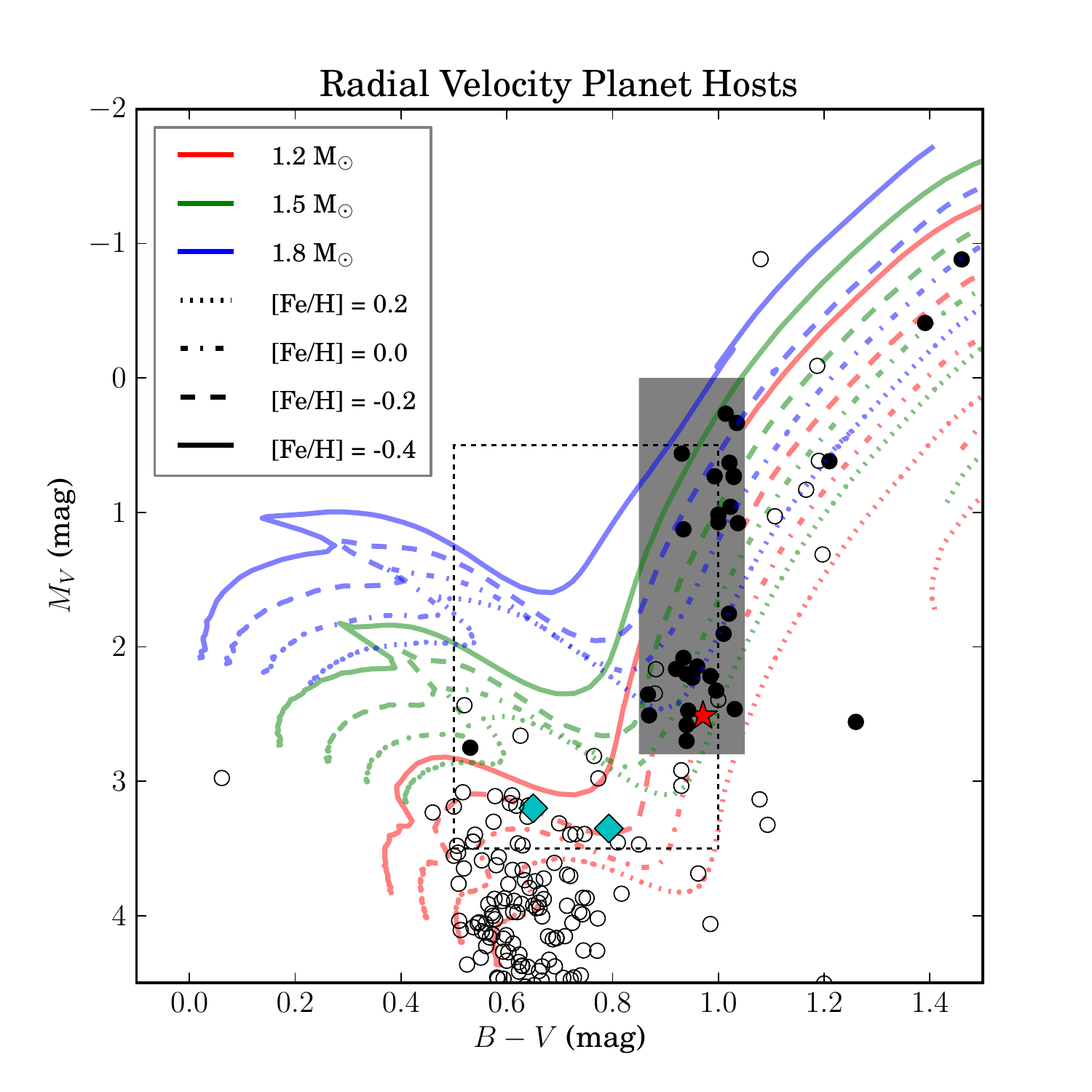}
\caption{Observational Hertzsprung-Russell diagram ($B-V$ color vs absolute $V$ magnitude) of host stars with planets as of May 2011.  Stars identified as $>1.5~M_\odot$ and $<1.5~M_\odot$ are shown as filled circles and open circles respectively.  The dotted line outlines the parameter space ($0.5<M_V<3.5;0.5<B-V<1.0$) of the sample of 159 evolved stars defined by \citet{Johnson:2006lr}.  Shaded is the region bounded by $0.0<\mathrm{M}_V<2.8;0.85<B-V<1.05$, inside of which 28 (of 35 total) planet hosts with mass $>1.5~M_\odot$ and only 3 with mass $<1.5~M_\odot$ reside.  The properties of this subsample and region are further described in Figure~\ref{fig:MDF} and Section~\ref{sec:constructhisto}.  In red, green and blue are evolutionary tracks derived from Yonsei-Yale isochrones \citep{Yi:2003fu} for 1.2, 1.5 and 1.8 M$_\odot$.  Line style indicates stpdf of metallicity from [Fe/H]=-0.4 to [Fe/H]=0.2.  }
\label{fig:BmVMv} 
\end{center}
\end{figure}

The discrepancy for two stars in Figure~\ref{fig:rotationstate}, HD 190228 and HD 154857, 
can be attributed directly to erroneous masses.  These mass identifications come from analysis of SPOCS catalog spectra \citep{Valenti:2005fk} with a Bayesian fit of atmospheric parameters to stellar evolution model grids \citep{Takeda:2007lr}.  In both cases, the mass likelihood function has two peaks and \citet{Takeda:2007lr} report secondary masses.  In the case of HD 190228 the adopted mass is $1.821 \pm0.042~M_\odot$, with an alternate mass of $1.119~M_\odot$ assigned 20\% probability.  In the case of HD154857, the adopted mass is $1.718^{+0.030}_{-0.022}~M_\odot$, and alternate mass $1.161~M_\odot$  assigned 59\% probability.   
There are alternate masses reported in the literature for 
HD 190228: $0.83~M_\odot$; 1.20$\pm0.05 ~M_\odot$; $0.96$--$1.51~M_\odot$ \citep{Perrier:2003lr,Allende-Prieto:1999yq,Valenti:2005fk} and 
HD 154857:
$1.17\pm0.05~M_\odot$; $1.35\pm0.12~M_\odot$ $0.98$--$1.62~M_\odot$ \citep{McCarthy:2004qy,Allende-Prieto:1999yq,Valenti:2005fk}.  High masses for these stars are inconsistent with their luminosity (see Figure~\ref{fig:BmVMv}).
The rotational discrepancy is therefore explained by incorrect mass determination alone.  With lower mass these stars are consistent with slowly rotating solar-type stars.

While it is not surprising that individual mass determinations may be erroneous, the question of the reliability of the masses is key to interpreting rotational discrepancies.  Comparison of the 25 evolved stars ($\log g>4$) in common between the Exoplanet Orbit Database and \citet{Allende-Prieto:1999yq} is shown in Figure~\ref{fig:EODvsAP99}.  These masses all rest in some way on placing observed properties on a model stellar evolution grid.  This is straightforward for main sequence stars, but increasingly problematic for evolved stars due to crossing of evolutionary tracks (see Figure~\ref{fig:BmVMv}).  With accurate determinations of composition and gravity from spectral analysis, the degeneracies can be broken.  However, the risk of erroneous mass estimates is substantial.  The determination of $\log~g$ by pressure broadening of the wings of the MgB triplet lines is less well determined for the low gravity (low density) regime of evolved stars, so the spectroscopic $\log~g$ determinations may include additional systematic errors not
accounted for in catalogs such as \citep{Valenti:2005fk}.
For all stars shown in Figure~\ref{fig:rotationstate}, it is possible to drastically change the mass determination with modest changes to the methods or uncertainties, even assuming there are no systematic uncertainties in the underlying stellar evolution models.  


While in general, stellar evolution codes have been refined to the point of high accuracy, there are regions of parameter space with known discrepancies.   At the base of the red giant branch for near solar mass stars, there are discrepancies as large as 0.5 dex in log L and 100K in $T_\mathrm{eff}$ with changes in equation of state or mixing length parameter from 1.6 to 1.9 \citep{Cassisi:2005qy}.  There are discrepancies between different stellar evolution models as large as 0.3 dex in Luminosity and 0.01 dex in log $T_{\mathrm{eff}}$ with changes in the treatment of equation of state and atmosphere optical boundary conditions \citep{Salaris:2002uq}.   Extrapolation of the solar-calibrated mixing length to convection zones at the base of the red giant branch may not be accurate, and lower gravity convection may introduce additional errors.  \citet{Robinson:2004fk} found that differences between mixing length theory and 3-D simulations increase as surface gravity decreases.  Applicability of mixing length theory to stars that are not quasi-stationary and undergoing the development of shells with superadiabatic convection is an area where the accuracy of the models can be called into question \citep{Cassisi:2010fj}.

The position of the red giant branch is dictated by the Hayashi line.  For $T_{\mathrm{eff}} \lesssim 5000$~K, the dominant opacity is H$^-$.  Since the H$^-$ opacity is proportional to the abundance of electrons from easily ionizable metals, changes in metallicity will change the opacity and move the Hayashi line.  A reduction of the metallicity in of 1.0 dex shifts the Hayashi line by 0.05 dex in log $T_{\mathrm{eff}}$, which changes the mass for a given $L$, $T_{\mathrm{eff}}$ by a factor of 2.  A change in the mixing length parameter $l_m/H_P$ from 1.5 to 1.0 will introduce an equivalent shift \citep{Henyey:1965pb}.  Changes in the $\alpha$-process abundance will similarly move the position of the red giant branch \citep{Kim:2002kx}. 

\begin{figure}[htbp]
\begin{center}
\includegraphics[width=6in]{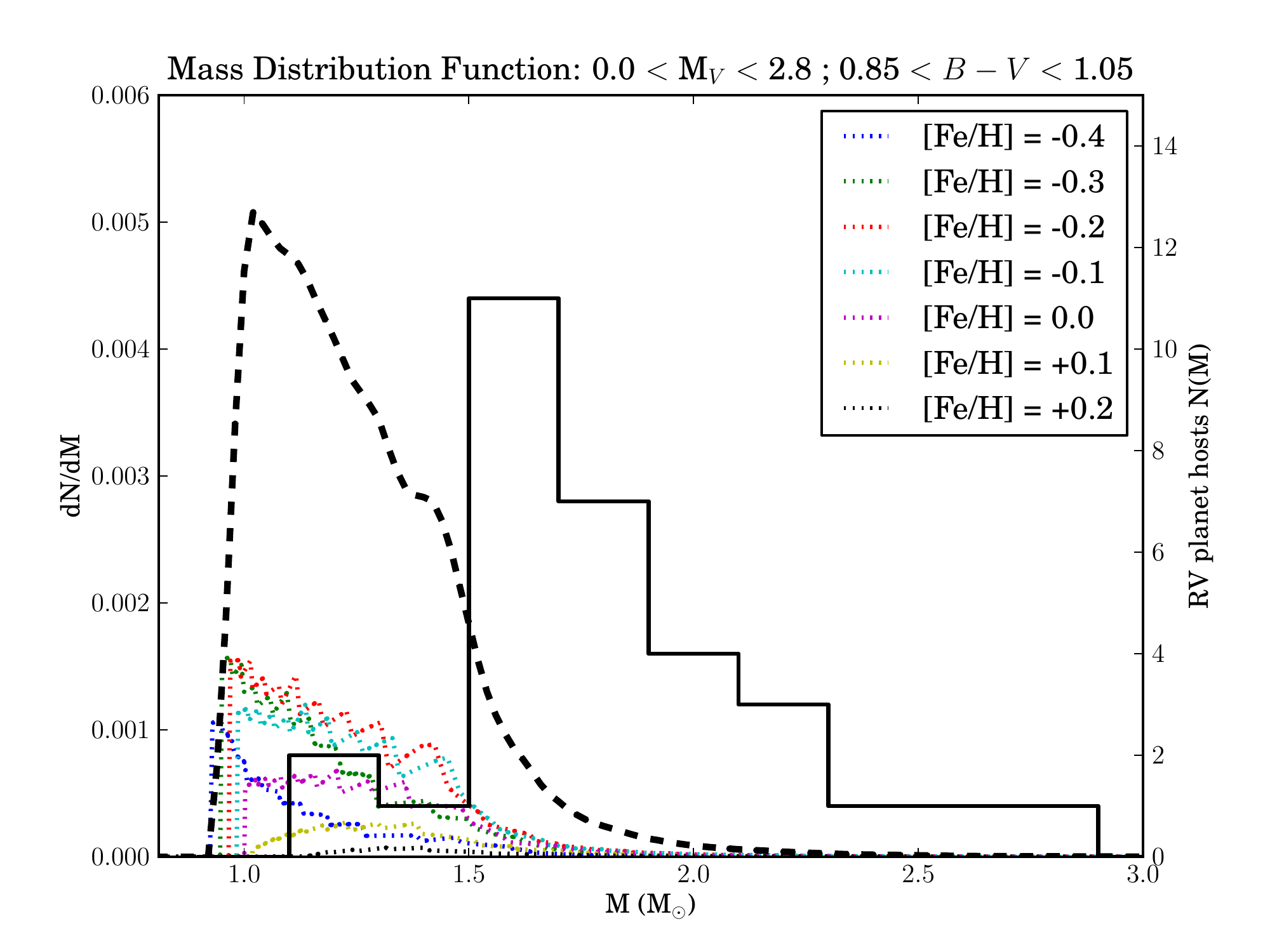}
\caption{Mass distribution function for exoplanet host stars in the shaded region of Figure~\ref{fig:BmVMv} ($0.0<\mathrm{M}_V<2.8;0.85<B-V<1.05$).  The dashed line is the distribution constructed by integrating isochrones as described in section~\ref{sec:constructhisto}.  Shown in dotted lines are the individual mass distributions for the metallicity grid.  The solid line is exoplanet host stars.  29 of 31 exoplanet host stars in this region of the HR diagram, but only 11\% of the integrated isochrone have mass $>1.5~M_\odot$.}
\label{fig:MDF} 
\end{center}
\end{figure}

The disagreement between the $\log g$ distributions of the $1.6$--$2.0~M_\odot$ samples in Figure~\ref{fig:rotationstate}, although suggestive that the host mass determinations may be erroneous, is not definitive. There are, however, aspects of stellar evolution that are very robust.  While the specific tracks change with parameters that are not well known, the rate a star evolves through the HR diagram is very insensitive.

\subsection{Mass Distribution}

\label{sec:constructhisto}

A statistical comparison of the mass distribution of evolved exoplanet host stars with the distribution of masses inferred from the rate at which stars evolve through the same region of the HR diagram is a robust test of the mass determination of the sample.     The expected mass distribution function for a region of the observational HR diagram can be constructed by integrating isochrones, weighted by an initial mass function, metallicity distribution and star formation history.  Figure~\ref{fig:MDF} shows such a mass distribution function, obtained by accumulating a synthetic mass distribution from the Yonsei-Yale isochrones weighted by the time stars of all masses spend in the region shaded in Figure~\ref{fig:BmVMv}, for a Salpeter IMF $dN/dM \propto M^{-2.35}$ \citep{Salpeter:1955uq}, uniform star formation history over 12 Gyr and gaussian distribution of metallicity with mean [Fe/H]=-0.14 and $\sigma$[Fe/H]=0.19 \citep{Nordstrom:2004fj}.  

The comparison between the mass distribution of evolved exoplanet host stars and the mass distribution expected from stellar evolution shown in Figure~\ref{fig:MDF} is striking.  The lack of massive stars in the expected mass distribution is primarily because stars more massive than $1.4~M_\odot$ ascend the giant branch more rapidly than less massive stars because the contraction of the hotter core is not slowed by degeneracy pressure. While it is possible that a dramatic enhancement of the planet frequency above $1.5~M_\odot$ accounts for the difference, it is much more likely that the mass inferences are incorrect, and the exoplanet hosts are in fact of lower mass.  In the absence of reliable masses, it is not possible to conclude anything concerning rotation state, but the reliability of the masses alone becomes the key question.

\section{Discussion}

If K giant stars hosting planets are not different in mass from the FGK dwarf stars hosting planets, then an alternate explanation for the differences in planet frequency and period distribution is needed.

A lack of short period planets is readily accounted for by tidal capture.  \citet{Villaver:2009uq} show $1 M_\mathrm{Jup}$ planets inside 3 AU and 2.1 AU are tidally captured during the red giant branch by stars of mass $1~M_\odot$ and $2~M_\odot$ respectively.   That the lower mass (counterintuitively) captures planets to a larger radius is a result of the slower evolution of the lower mass star.  This accretion {\em should} show signatures in the angular momentum of the planet hosts, but since the angular momentum evolution of a star changes drastically with mass, more reliable masses are needed to perform that test.

The increased planet frequency is more difficult to explain.  Tidal evolution could concentrate planets in a detectable range of periods, where they would not be detectable in dwarfs.  An age-metallicity relation would bias the giants to be metal-poor, consistent with the distribution found by \citep{Pasquini:2007zr}, leading to the expectation that a K giant population should be planet-poor on the basis of the planet-metallicity correlation.  The giant branch can be expected to be enhanced in metal-rich stars, since these stars evolve more slowly (see Figure~\ref{fig:MESA}).  It is possible that the giant metallicities are systematically in error, and the population is metal rich, but to achieve the observed enhancement in planet abundance such a scenario would require a recent episode of super-metal-rich star formation for which there is no other evidence.   It is possible that the accretion of planets alters the atmosphere or stellar structure in a way that concentrates the planet-hosting giants in a particular region of the HR-diagram, altering the relative frequency of planet-bearing and non-planet-bearing stars.  Such explanations are, however, difficult to contrive to enhance the planet abundance by the observed factor of at least two between the FGK dwarf sample and giant sample, and do not account for the abrupt jump at $1.5~M_\odot$ in Figure~\ref{fig:MDF}.  The least speculative interpretation is that the increase in planet abundance is due to a substantial contamination in the K giant planet sample from non-planetary signals due to stellar variability.  Line bisector analyses have not shown evidence of stellar origin for these signals, although it is possible for bisector variations to be too small to be detected, yet still account for the radial velocity signal (e.g. \citet{Reffert:2006lr}).

Definitive tests of stellar mass for the K giant planet hosts and methods to confirm the planets are needed.  In the case of stars that reside in binary systems, it is possible to determine dynamical masses, but systems that are in a configuration to do so are rare.  Asteroseismology can accurately determine the physical parameters of stars, with space-quality photometry.  \citep{Kallinger:2010lr} determined the masses of 1000 giants with Kepler, and found a similar mass distribution to the synthetic mass distribution in Figure~\ref{fig:MDF}.

\subsection{Possible Tests}

\subsubsection{Transits}

The large size of giant stars is advantageous for confirmation of planets by transits because of the increase in transit probability $\sim R_*/a$.  
\citet{Kane:2010lr} notes the high likelihood of transits of $\iota$ Dra and HD 122430.  
Since $a$ is inferred from Kepler's law $a^3\propto M P^2$, the transit probability (for a stellar radius determined independently of stellar mass) is proportional to $M^{-1/3}$, so revising the stellar masses down increases transit probability.  A key test case would be the 6 day period planet orbiting HD 102956, which has $R_*/a = 0.27$, even for an assumed mass of 1.7 $M_\odot$.  Unfortunately the large stellar radius dilutes transit depth making these challenging observations, but transits in limb-brightened emission lines are promising \citep{Schlawin:2010ly}.  Radii of can be accurately measured by interferometry \citep{van-Belle:2009ve,Baines:2010gf}, which combined with stellar density determined from the transit lightcurve \citep{Seager:2003ul} provides a model independent mass.

\subsubsection{Isotopes Ratios}

\begin{figure}[htbp]
\begin{center}
\includegraphics[width=4.5in]{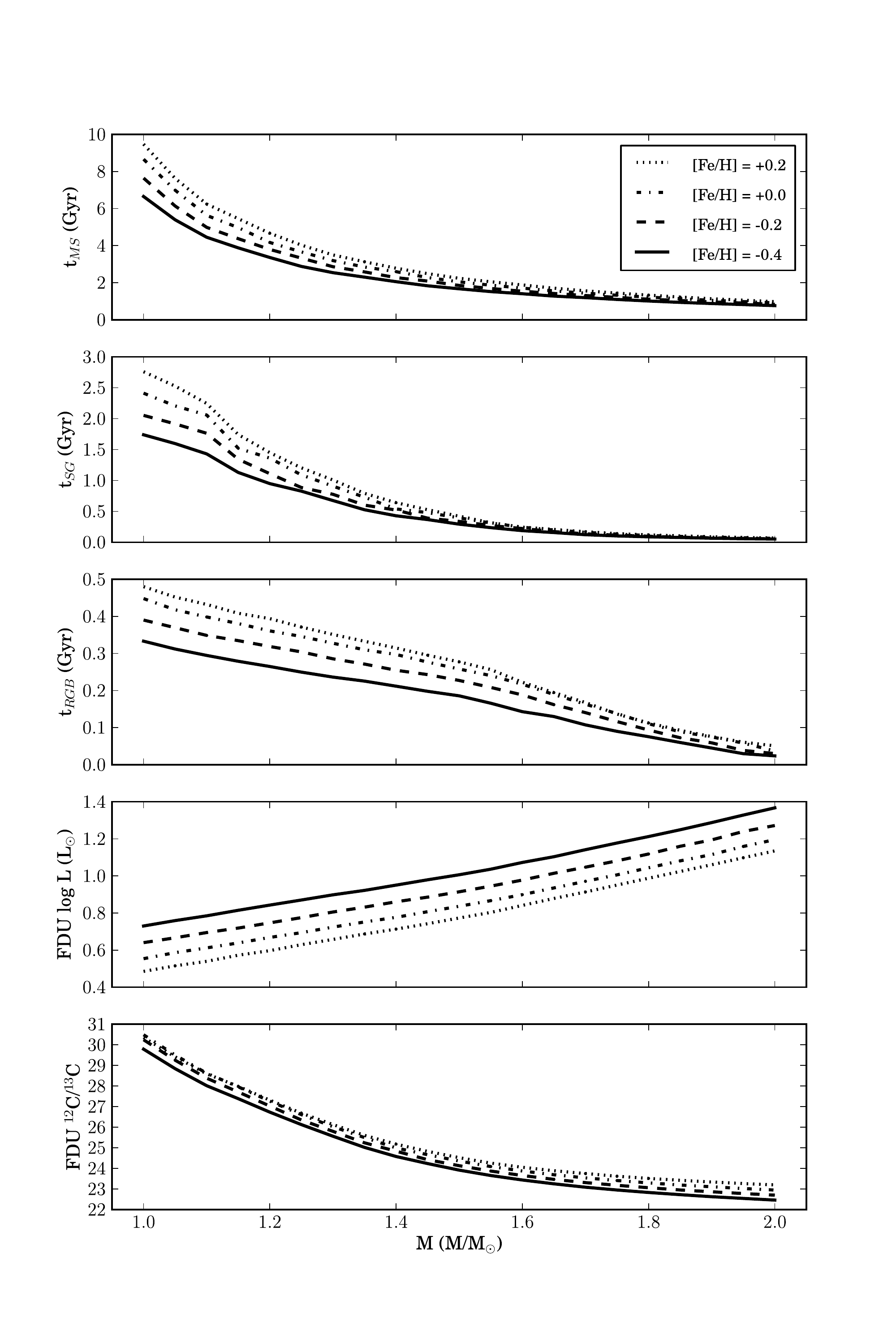}
\caption{MESA \citep{Paxton:2011pd} stellar evolution calculations, showing sequentially: main sequence lifetime; Hertzsprung Gap crossing time ($t_{HG}$, defined as the time between core Hydrogen exhaustion and first dredge-up): red giant branch lifetime ($t_{RGB}$, defined as the time between first dredge-up and the $L=100~L_\odot$); the luminosity at which first dredge-up occurs; post dredge-up $^{12}$C/$^{13}$C ratio. The calculations are in good agreement with the $^{12}$C/$^{13}$C ratios from \citep{Eggleton:2008lr}.}
\label{fig:MESA} 
\end{center}
\end{figure}

It would be highly desirable to determine stellar masses of planet hosting K giants in a robust and uniform manner.  A possible approach is the measurement of atmospheric Carbon isotope ratios.  As a star evolves into a giant, the development of a thick convective envelope brings products of nuclear burning to the surface, changing the photospheric abundance of elements that participate in the nuclear burning such as $^{3}$He, $^{13}$C, $^{14}$N, $^{17}$O and $^{18}$O.  The equilibrium isotope ratios of the CNO elements are sensitive to the temperature of the CNO cycle, so these isotope ratios are sensitive to the stellar mass.  Although the mixing processes are not well known for very evolved giants, the $^{12}$C/$^{13}$C ratio through the first dredge-up is well understood and theory and observations are in agreement \citep{Charbonnel:1994qf}.   The $^{12}$C/$^{13}$C ratio drops to a range of  30 to 20 in first dredge-up (FDU) giants of 1 to 2 $M_\odot$, with weak dependance on metallicity and very little dependance on mixing length parameter \citep{Charbonnel:1994qf}.  Figure~\ref{fig:MESA} shows the photospheric $^{12}$C/$^{13}$C calculated with the MESA stellar evolution code \citep{Paxton:2011pd} for post-FDU giants as a function of mass and metallicity.
The $^{17}$O/$^{18}$O ratio is potentially a sensitive diagnostic, but the reaction rates are less well known and the observations less well in accord with theory \citep{Stoesz:2003bh}.

The $^{12}$C/$^{13}$C ratio can be used to place a star convincingly as a Hertzprung Gap star, since the first dredgeup occurs at the base of the RGB and the $^{12}$C/$^{13}$C remains near the primordial (solar value of 80 \citep{Ayres:2006lq}) until the star joins the RGB.  Most of the K giant planet hosts have a luminosity above the first dredge-up, but HD 192699 ($M=1.7\pm0.12~M_\odot , T_{eff}=5220\pm44~\mathrm{K}, L=1.04~L_\odot, \mathrm{[Fe/H]}=-0.15\pm0.04$) could be before or after FDU within the mass uncertainty.

\acknowledgements

This research has made use of the Exoplanet Orbit Database at exoplanets.org and NASA's ADS Bibliographic Services. This work has been partially supported by the National Science Foundation under Grant No. AST-0905932.  I thank  Lars Bildsten, Kevin Covey, Dan Fabrycky, Mike Ireland, Adam Kraus, Jeff Valenti and Jason Wright for discussions and Bill Paxton for help with the MESA evolutionary code.

\end{document}